\documentclass[%
reprint, twocolumn,
superscriptaddress,
showpacs,preprintnumbers,
footinbib,
amsmath,amssymb, aps, prl,
]{revtex4}

\usepackage{graphicx}
\usepackage{dcolumn}
\usepackage{bm}
\usepackage{color} 
\usepackage{CJK}
\usepackage{dsfont}

\def\la{{\langle}}
\def\ra{{\rangle}}

\newcommand{\beq}{\begin{equation}}
\newcommand{\eeq}{\end{equation}}
\newcommand{\beqa}{\begin{eqnarray}}
\newcommand{\eeqa}{\end{eqnarray}}

\begin{document}
\title{Fast quasi-adiabatic dynamics}
\author{S. Mart\'\i nez-Garaot}
\affiliation{Departamento de Qu\'{\i}mica F\'{\i}sica, UPV/EHU, Apdo.
644, 48080 Bilbao, Spain}
\author{A. Ruschhaupt}
\affiliation{Department of Physics, University College Cork, Cork, Ireland}
\author{J. Gillet}
\affiliation{OIST Graduate University, Onna, Okinawa 904-0411, Japan}
\author{Th. Busch}
\affiliation{OIST Graduate University, Onna, Okinawa 904-0411, Japan}
\author{J. G. Muga}
\affiliation{Departamento de Qu\'{\i}mica F\'{\i}sica, UPV/EHU, Apdo.
644, 48080 Bilbao, Spain}
\affiliation{Department of Physics, Shanghai University, 200444
Shanghai, People's Republic of China}
\date{\today}
\begin{abstract}
We work out the theory and applications 
of a fast quasi-adiabatic approach to speed up slow adiabatic
manipulations of quantum systems  by driving a control parameter
as near to the adiabatic limit as possible over the entire protocol duration. 
Specifically, we show that the population inversion in a two-level system, the splitting and cotunneling 
of two-interacting bosons, and the stirring of a Tonks-Girardeau gas on a ring to achieve mesoscopic superpositions
of many-body rotating and non-rotating states, can be significantly speeded up.    
\end{abstract}
%
%
\maketitle
%
%
%
%
Developing  technologies based on delicate quantum coherences of  
atomic systems is a major scientific and technical challenge due to pervasive noise-induced and manipulation errors.  
Shortening the process  below characteristic decoherence times provides
a way out to avoid the effects of noise, but the protocol (time dependence of control parameters) 
should still be robust with respect to offsets of the external driving parameters. 
Shortcuts to adiabaticity (STA) are a set of techniques to reduce the duration of slow adiabatic processes, 
minimizing noise effects while keeping or enhancing robustness \cite{Chen_PRL2010,noise,review}.  
There are  different approaches but they are not always easy to implement in practice, because of 
the need to control many variables, or the difficulty to realize certain
terms added to the original  Hamiltonian (the ones that allow the adiabatic dynamics to be sped up). 
Here we  work out the theory and present several applications of a simple, but effective, fast quasi-adiabatic (FAQUAD) approach that  
engineers the time dependence of a single control parameter $\lambda(t)$, 
without changing the structure of the original Hamiltonian,  $H[\lambda(t)]$,  
to perform a process as quickly as possible while making 
it as adiabatic as possible at all times. 
The two goals are contradictory so a compromise is needed. 
We impose that the standard adiabaticity parameter \cite{schiff} is constant throughout
the process, and consistent with 
the boundary conditions  (BC) of $\lambda(t)$ at $t=0$ and $t=t_f$.   

In the simplest scenario we assume that the adiabatic process driven by changing $\lambda(t)$ involves a passage through at least one avoided crossing.
While real systems are in general multilevel, only the two quasi-crossing levels (say $E_1$, $E_2$) in the
instantaneous basis $\{|\phi_j\ra\}$ need to be considered under the adiabaticity condition \cite{schiff}, 
$\hbar \left |\frac{ \langle \phi_1(t)|\partial_t \phi_2(t)\rangle}{ E_1(t)-E_2(t)} \right |\ll1$. 
(More levels can be taken into account if necessary.)  
We then impose 
\beq
\label{f_adiabatic}
\hbar \left |\frac{ \langle \phi_1(t)|\partial_t \phi_2(t)\rangle}{ E_1(t)-E_2(t)} \right |=
\hbar \left |\frac{ \langle \phi_1(t)|\frac{\partial H}{\partial t}| \phi_2(t)\rangle}{ [E_1(t)-E_2(t)]^2} \right | =c,
\eeq 
and as $\lambda=\lambda(t)$ and $t=t(\lambda)$ we apply the chain rule to write    
\beq
\label{d_e}
\dot \lambda\!=\!\mp\frac{c}{\hbar}Ê\!\left | \frac{ E_1(\lambda)-E_2(\lambda)}{ \langle \phi_1(\lambda)|\partial_\lambda \phi_2(\lambda)\rangle} \right |\!
\!=\!\mp\frac{c}{\hbar}\!\left | \frac{ [E_1(\lambda)-E_2(\lambda)]^2}{ \langle \phi_1(\lambda)|\frac{\partial H}{\partial \lambda}|\phi_2(\lambda)\rangle} \right |\!,
\eeq
where the overdot is a time derivative and $\mp$ applies to a monotonous decrease/increase of $\lambda(t)$.  
Eq. (\ref{d_e}) must be solved with the BC 
$\lambda(0)$ and  $\lambda(t_f)$, which fixes $c$ and the integration constant. 
The corresponding fast quasi-adiabatic (FAQUAD) solution, $\lambda_{F}(t)$,
changes fast when the transitions are unlikely and slowly otherwise. 
An equation equivalent to Eq. (\ref{d_e}) has been applied to specific
models \cite{Phillips,Guerin07,Guerin08,fa,Bowler,multi}, for example
the two-level system \cite{Guerin08} and three-level lambda system
\cite{Guerin07}.


In this paper, we derive 
important properties of FAQUAD including characteristic time
scales, such as the minimal time to achieve fidelity one, and its optimality within the iterative superadiabatic sequence. 
We  apply FAQUAD to several physical systems for which other shortcut techniques are difficult or impossible
to implement, 
including a process for creating a collective superposition state between rotating
and non-rotating atoms on a ring.

The FAQUAD strategy belongs to a family of processes that use the
time-dependence of a control parameter to delocalize  in time the transition
probability among adiabatic levels. 
%
In the parallel adiabatic transfer technique \cite{Guerin,Guerin1} the level
gap is required to be constant, which prevents it from being applicable when 
the initial and final gaps are different (see the TG gas
example below). The uniform adiabatic (UA) method developed in \cite{Zurek}
relies on a comparison of {\it transition} and {\it relaxation} time scales
and predicts (in a notation consistent with the one used in the work)
%
%
$
\dot \lambda
=\mp\frac{c_{UA}}{\hbar}\left | \frac{ [E_1(\lambda)-E_2(\lambda)]^2}{\partial [E_1(\lambda)-E_2(\lambda)]/\partial \lambda} \right |.
$
%
%
Furthermore, the local adiabaticity (LA) approach \cite{Cerf,Monroe} predicts an equation similar to Eq.~(\ref{d_e}), however without the factor $\langle \phi_1(\lambda)|\frac{\partial H}{\partial \lambda}| \phi_2(\lambda)\rangle$. This leads to a different constant, $c_{LA}$, and 
time dependence of the parameter, $\lambda_{LA}(t)$, and therefore different minimal times as illustrated below.      
Note that in \cite{Cerf}, Eq. (\ref{d_e}) is also written down but not applied
as such.




%
{\it{General Properties.}}
We rewrite Eq. (\ref{d_e}) in terms of  $s=t/t_f$ and we define $\tilde{\lambda}(s):=\lambda(s t_f)$ so that  
$
\dot \lambda(t)=\tilde{\lambda}' \frac{1}{t_f}, 
$
where the prime is the derivative with respect to $s$. We get
\beq
\label{deltap}
\tilde{\lambda}'=\mp\frac{\tilde{c}}{\hbar} \left|  \frac{{E}_1-{E}_2}{ \langle {\phi}_1|\partial_{\tilde{\lambda}}\phi_2\rangle} \right |_{\tilde{\lambda}}, 
\eeq
with $\tilde{c}=c t_f=\mp\hbar\int_{\tiny{\tilde{\lambda}(0)}}^{\tiny{\tilde{\lambda}(1)}} d{\tilde{\lambda}}/ \big|  \frac{{E}_1-{E}_2}{ \langle {\phi}_1|\partial_{\tilde{\lambda}}\phi_2\rangle} \big|_{\tilde{\lambda}}$. 
It is thus enough to solve the FAQUAD protocol once, i.e. using Eq. (\ref{deltap}) we get $\tilde{\lambda}_F(s)$ and $\tilde{c}$ to satisfy  
$\tilde{\lambda}(s=0)$ and $\tilde{\lambda}(s=1)$, and then 
adapt (scale) the result for each $t_f$, as $\lambda_F(t=s t_f)=\tilde{\lambda}_F(s)$, and $c=\tilde{c}/t_f$. 
Similarly, the gap $\omega_{12}(t)=[E_1(t)-E_2(t)]/\hbar$ is given in terms of a universal gap function $\tilde{\omega}_{12}[\tilde{\lambda}_F(s)]$ as 
$\omega_{12}(t)=\tilde{\omega}_{12}[\tilde{\lambda}_F(t/t_f)]$.   
Depending on $\tilde{c}$, a large time $t_f$ 
might be necessary to make the process fully adiabatic (i.e., with a small enough $c$) but, surprisingly, much shorter times for which the 
process is not fully adiabatic also lead to the desired results.      
Since the system is nearly adiabatic this is explained by adiabatic perturbation theory.    
In the adiabatic basis the wave function is expanded as \cite{schiff,ac_non-Hermitian}
$
|\Psi(t)\rangle=\sum_n g_n(t) e^{i\beta_n (t)}|\phi_n(t)\rangle,
$
where $\beta_n(t)=-\frac{1}{\hbar}\int_0^t E_n(t')dt'+i\int_0^t\langle\phi_n(t')|\dot \phi_n(t')\rangle dt'$.
From  $i\hbar|\dot \Psi(t)\rangle=H_0(t)|\Psi(t)\rangle$ we get, choosing
$\langle\phi_n(t)|\dot \phi_k(t)\rangle$ to be real (in particular $\langle\phi_n(t)|\dot \phi_n(t)\rangle=0$), 
%
$
\dot g_n(t)=-\sum_{k\neq n} e^{iW_{nk}(t)}\langle\phi_n(t)|\dot \phi_k(t)\rangle g_k(t),
$
%
where $W_{nk}(t)=\int_0^t \omega_{nk}(t')dt'$ is a dynamical-gap phase and $\omega_{nk}(t):=[E_n(t)-E_k(t)]/\hbar$.
Integrating, 
$
g_n(t)\!-\!g_n(0)\!=\! \!-\!\sum_{k\neq n}\!\int_0^t\!\!e^{iW_{nk}(t')}\!\langle\phi_n(t')|\dot \phi_k(t')\rangle g_k(t') dt',
$
which is still exact. 
Assuming that the initial state is $|\phi_m(0)\rangle$ and approximating $g_k(t')=\delta_{km}$
one finds to first order, for $n\neq m$,
%
$
g_n^{(1)}(t)=-\int_0^t\langle\phi_n(t')|\dot \phi_m(t')\rangle e^{iW_{nm}(t')}dt',
$
%
which should satisfy $|g_n(t)|\ll1$ for an adiabatic evolution.
In FAQUAD
$
\langle\phi_n(t)|\dot \phi_m(t)\rangle=c r\omega_{nm}(t),
$
with 
$
r={\rm{sgn}}[\langle\phi_n(t)|\dot \phi_m(t)\rangle \omega_{nm}], 
$
so we find (higher order corrections are also explicit) 
\beqa
g_n^{(1)}(t)\!=\!-{r}\!\int_0^t\!\!\! c \omega_{nm}(t') e^{iW_{nm}(t')}dt' 
\!=\!ic{r}
(e^{iW_{nm}(t)}\!-\!1).
\label{fa_perturbation_sol}
\eeqa
Note the scaling  $W_{nm}(t_f)=t_f \Phi_{nm}$ where $\Phi_{nm}=\int_0^1 \tilde{\omega}_{nm}(s) ds$, 
and $\tilde{\omega}_{nm}(s)=\omega_{nm}(s t_f)$.    
The oscillation period for the final population with FAQUAD
is 
$
T=\frac{2\pi}{\Phi_{12}}, 
$
which is also a good estimate of the minimal time for a complete transition. 
The upper envelope for the $n$-th state probability is $4\tilde{c}^2/t_f^2$.  
The period, envelope, and Eq. (\ref{fa_perturbation_sol}) are important general results
of this work.  
The oscillation is due to a   
quantum interference:   
$g_n^{(1)}(t_f)$ results from the sum of paths where 
the jump at time $t'$ from $m$ to $n$ has an amplitude  $c\omega_{nm}(t')$.  
$e^{iW_{nm}(t')}$ represents the dynamical phases before and after the jump, as $
e^{iW_{nm}(t')}= e^{\frac{-i}{\hbar}\!\int_0^{t'}\!dt''E_m(t'')} e^{\frac{-i}{\hbar}\!\int_{t'}^{t_f}\!dt''E_n(t'')} e^{\frac{i}{\hbar}\!\int_{0}^{t_f}\!dt''E_n(t'')}$, 
where the last exponential is a phase factor independent of $t'$.  

To illustrate these general properties, we will first examine the two-level model, a
paradigmatic testbed. Then, to show the power of FAQUAD,
we will apply it to more complicated atomic systems.


%
{\it{Population inversion.}}
Consider first a  two-mode model
with a single avoided crossing in the light of these time scales.
In the bare basis, $|1\rangle=\left (\scriptsize{\begin{array} {rccl} 1\\0 \end{array}}\right)$ and $|2\rangle=\left ( \scriptsize{\begin{array} {rccl} 0\\1 \end{array}} \right)$,  
the time-dependent state is 
$
|\Psi(t)\rangle=b_1(t)|1\rangle+b_2(t)|2\rangle
$
and 
$
H_0=\left(\scriptsize{\begin{array}{cc}
0 & -\sqrt{2}J
\\
-\sqrt{2}J & U-\Delta
\end{array}} \right),
$
%
where the bias $\Delta=\Delta(t)$ is the control parameter, and $U>0$, $J>0$, are constant.  
The goal is to drive the eigenstate from $|\phi_1(0)\rangle=|2\rangle$ to $|\phi_1(t_f)\rangle=|1\rangle$. To design the reference adiabatic protocol 
we impose on $\Delta(t)$ the BC
$\Delta(0)\gg U,J$ and $\Delta(t_f) =0$.
%
The FAQUAD protocol is shown in Fig. 1 (a) compared to a linear-in-time $\Delta(t)$ and a 
constant $\Delta=U$.  
%
%
%
%
%
%
%
%
\begin{figure}[t]
\begin{center}
\includegraphics[height=2.98cm,angle=0]{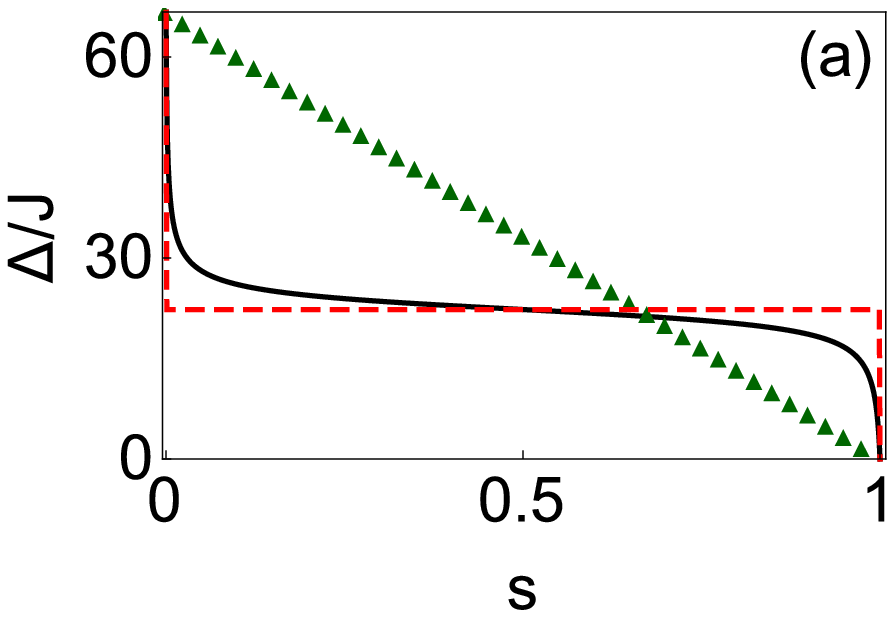}
\includegraphics[height=2.98cm,angle=0]{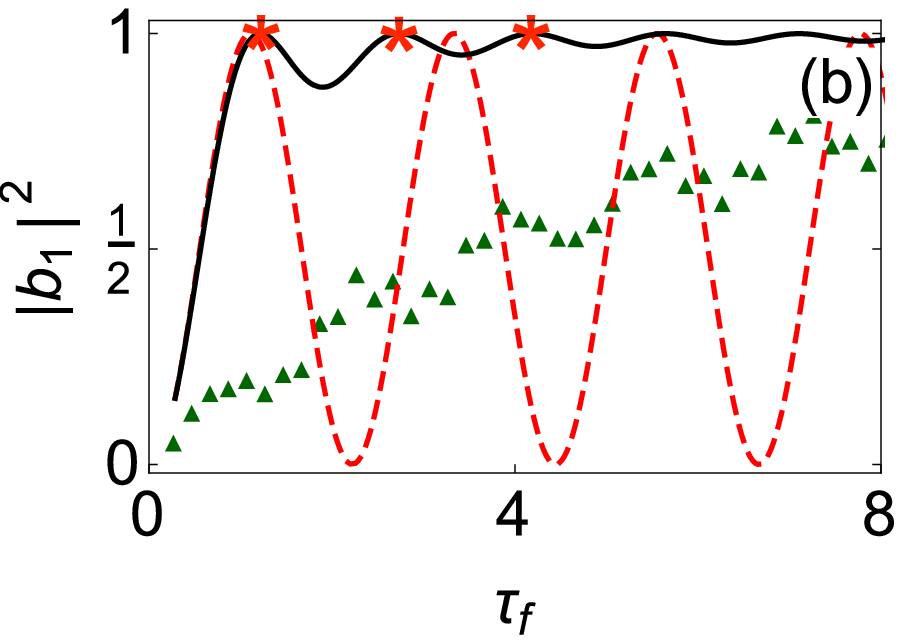}
\includegraphics[height=2.86cm,angle=0]{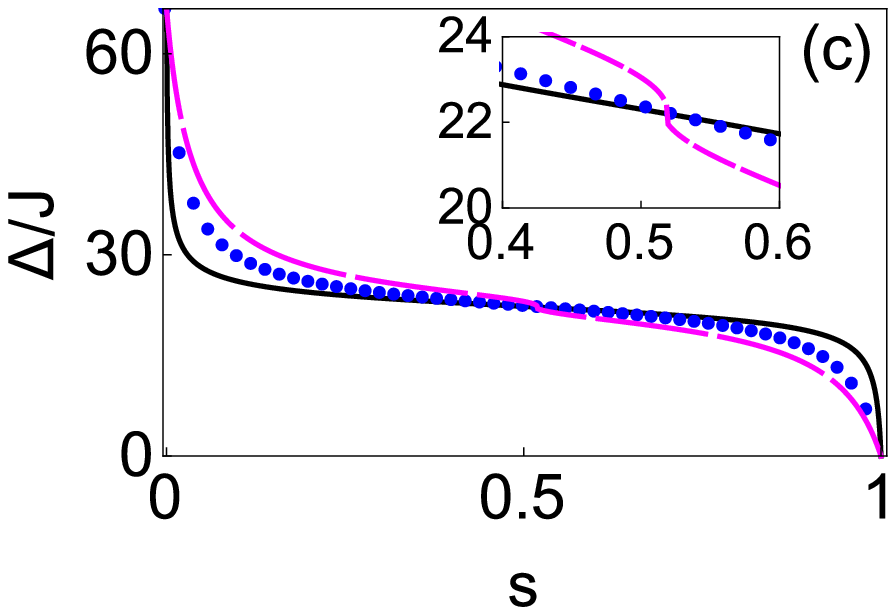}
\includegraphics[height=2.86cm,angle=0]{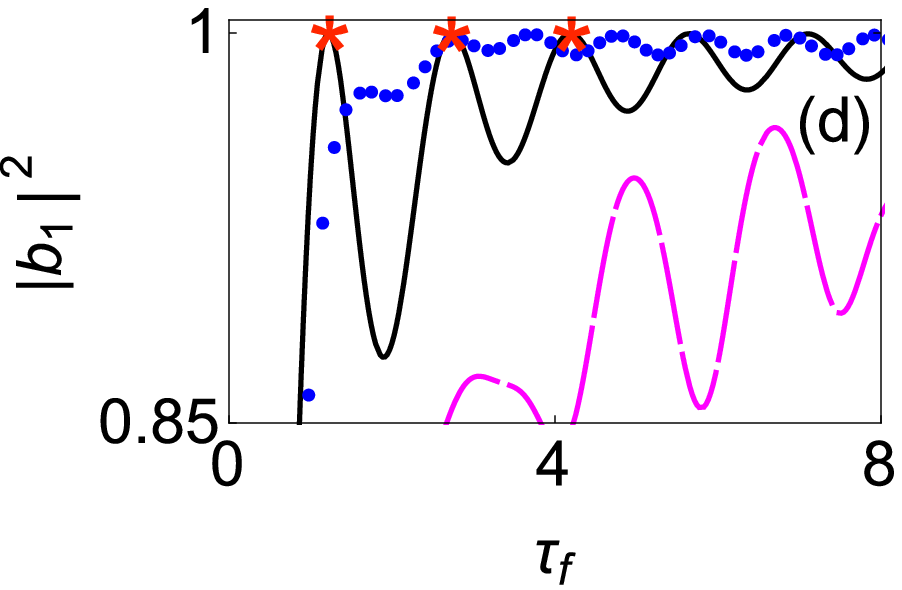}
\end{center}
\caption{\label{finalt_tmm}
(Color online). (a) Bias versus $s$ for linear-in-time bias (green triangles), $\pi$-pulse (short-dashed red line), and 
FAQUAD (solid black line).  
(b) Final ground state population $|b_1(t_f)|^2$ vs.   
$\tau_f=Jt_f/\hbar$ for linear-in-time bias (green triangles), $\pi$-pulse
(short-dashed red line), and FAQUAD (solid black line).
(c) Bias vs. $s$ for FAQUAD (solid black line), LA approach (blue dots), and UA approach  (long-dashed magenta line).  
The inset amplifies the kink of the UA approach.
(d) $|b_1(t_f)|^2$ vs. $\tau_f=Jt_f/\hbar$ for FAQUAD (solid black line), 
LA approach (blue dots), and UA approach  (long-dashed magenta line). The
stars in (b) and (d) correspond to integer multiples of the characteristic
FAQUAD time scale $2\pi/\Phi_{12}$.
$\Delta(0)/J=66.7$, $U/J=22.3$.}
\end{figure}
%
%
%
%
%
%
The final ground state populations $|b_1(t_f)|^2$ versus dimensionless final time $\tau_f=Jt_f/\hbar$ are shown in Fig. \ref{finalt_tmm} (b).
Since the dressed states are essentially pure bare states at initial and final times their populations in bare and dressed state bases coincide at these times.  
For $\Delta=U$ between 0 and $t_f$, 
Rabi oscillations occur (see Fig. \ref{finalt_tmm} (b)). The conditions for a $\pi$-pulse 
or multiple $\pi$-pulses are met periodically over $t_f$ alternated with times 
where the probability drops to zero because of 
destructive interference among two dressed states superposed with equal weights. 
By contrast the 
FAQUAD process is dominated by one dressed state and the influence of the transitions to the other one is minimized, because they are small in amplitude, and because at certain times they completely cancel each other out by 
destructive interference. The time interval between population maxima for
FAQUAD is $2\pi/\Phi_{1,2}$ (also shown in Fig. \ref{finalt_tmm} (b) and (d), stars),
i.e., it is not governed by the Rabi frequency. The first maximum is at a small $t_f$ similar to the one for the $\pi$-pulse, but broader. 
The FAQUAD maxima are more stable with respect to errors in $\Delta$ as $t_f$ increases, whereas the flat-pulse maxima decrease their stability.  
Fig. 1 (b) also shows the poorer results of the linear ramp for $\Delta(t)$.


%
%
%
%
%
%
%

FAQUAD is compared to the LA and UA approaches in Fig. \ref{finalt_tmm}(c) and
(d). It provides  shortcuts at smaller process times (it achieves $0.9998$ 
probability three times faster than LA) 
and an analytically predictable behavior
via the perturbation theory analysis.  
Let us now consider more complicated atomic systems where FAQUAD can be applied whereas other 
STA techniques cannot.  
%
%
%
%
%
%

\begin{figure}[t]
\begin{center}
(a)\includegraphics[height=1.06cm,angle=0]{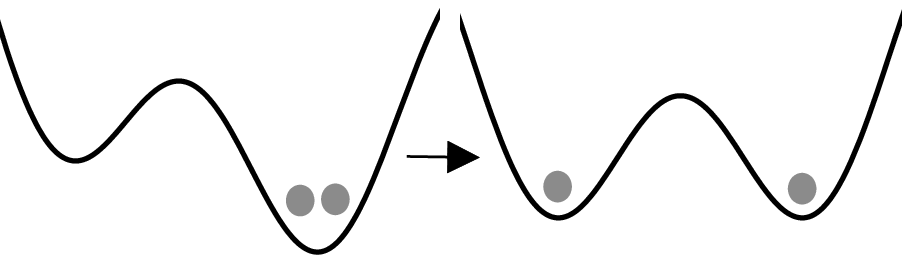}
\hspace{0.6cm}
(b)\includegraphics[height=1.06cm,angle=0]{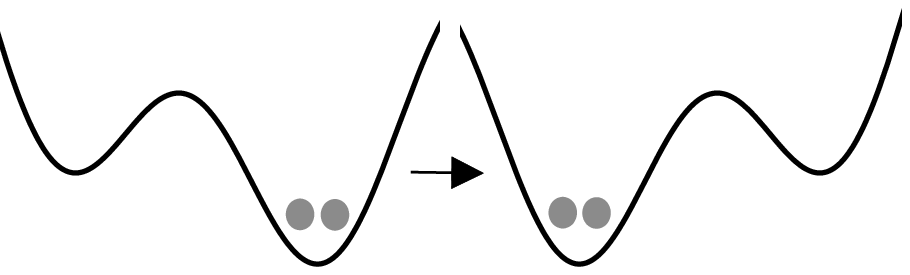}
\end{center}
\caption{\label{tunneling}
(a) Schematic representation of splitting from $|0,2\ra$ to $|1,1\ra$. (b) Cotunneling from
$|0,2\ra$ to $|2,0\ra$.}
\end{figure}

{\it{Interacting bosons in a double well.}}
Pairs of interacting bosons in a double well potential may be manipulated to implement universal quantum logic gates for quantum computation 
or to observe fundamental phenomena such as cotunneling  of two atoms
\cite{Phillips1,nature_Bloch}.
We shall speed up two processes: the splitting of the two particles from one to the two separate wells, 
and cotunneling, see Fig. \ref{tunneling}. 
The boson dynamics in a double well with tight lateral confinement 
is described by a two-site Bose-Hubbard Hamiltonian \cite{nature_Bloch}.
The Hamiltonian in the occupation number basis $|2,0\rangle=\left (\scriptsize{\begin{array} {rcccl} 1\\ 0\\0 \end{array}} \right)$, $|1,1\rangle=\left ( \scriptsize{\begin{array} {rcccl} 0\\ 1\\0 \end{array}} \right)$ and $|0,2\rangle=\left ( \scriptsize{\begin{array} {rcccl} 0\\ 0\\1 \end{array}} \right)$, is   
%
$\scriptsize{
H_0=\left ( \begin{array}{ccc}
U+\Delta & -\sqrt{2}J & 0\\
-\sqrt{2}J & 0 & -\sqrt{2}J \\
0 & -\sqrt{2}J & U-\Delta
\end{array} \right)},
$
where the bias $\Delta=\Delta(t)$ is the control function, $J$ is the hopping energy and $U$ the interaction energy. 
We write the time-dependent states as 
$|\Psi(t)\rangle=c_1(t)|2,0\rangle+c_2(t)|1,1\rangle+c_3(t)|0,2\rangle$.
%
%
%
%
%
%
%
%
%
%
%
%
%
%
%
Adiabatic processes that change $\Delta(t)$ slowly, keeping the $U/J$ ratio constant
are possible to implement splitting or cotunelling. 
Speeding them up by a ``counterdiabatic'' approach  is 
not possible in practice because of the need to apply new terms in the Hamiltonian which are difficult to implement. 
Alternative techniques could not be applied \cite{Molmer} or are  
cumbersome \cite{inv_algebra,sta_lie} because of the relatively large algebra involved.  
The FAQUAD approach provides a viable way out.  

%
%
%
%
%
%
%
%
\begin{figure}[t]
\begin{center}
\includegraphics[height=2.86cm,angle=0]{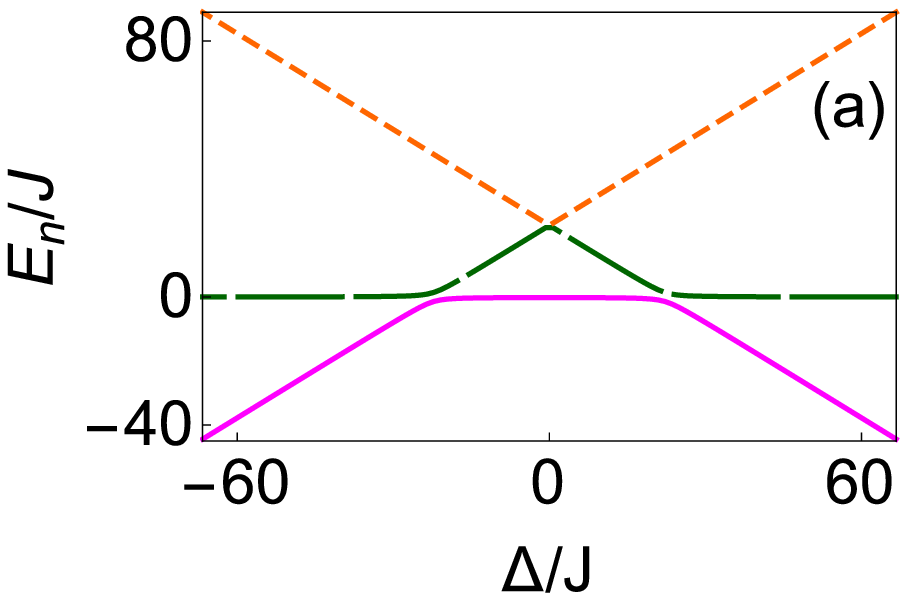}
\includegraphics[height=2.86cm,angle=0]{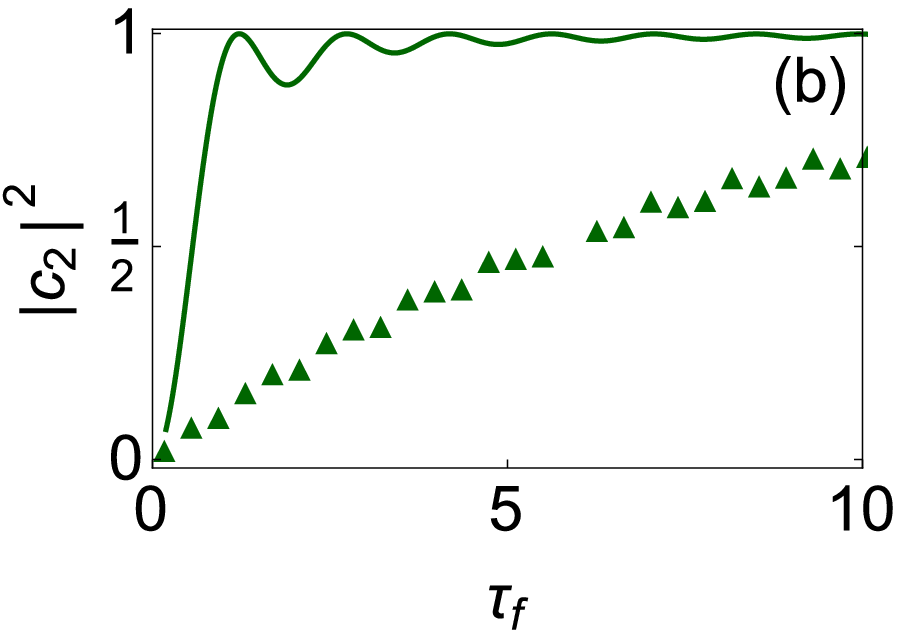}
\end{center}
\caption{\label{levels_ap}
(Color online)
(a) Energy levels vs. $\Delta$. For $n=1,2,3$: $E_1$ (solid magenta line), $E_2$ (long-dashed green line)
and $E_3$ (short-dashed orange line).  
$U/J= 22.3$. (b) $|c_2|^2$ vs. $\tau_f$ for linear-in-time bias (green triangles) and FAQUAD (solid green line). 
$\Delta(0)/J=100$, $U/J=33.45$, and $\tau_f=Jt_f/\hbar$.}
\end{figure}
%
%
%
%
%
%
- In a splitting process   $\Delta(0)\gg U,J$ and  $\Delta(t_f) =0$, 
see Fig. \ref{tunneling} (a).  
The initial ground state is $|\phi_1\rangle=|0,2\rangle$ and the final ground state $|\phi_1\rangle=|1,1\rangle$. 
Figure \ref{levels_ap} (a) shows the dependence of the three eigenenergies
with $\Delta$. 
$\Delta_{F}(t)$ is very similar to the result for the two-level system in Fig \ref{finalt_tmm} (a). 
%
%
%
%
%
%
%
%
\begin{figure}[t]
\begin{center}
\includegraphics[height=2.86cm,angle=0]{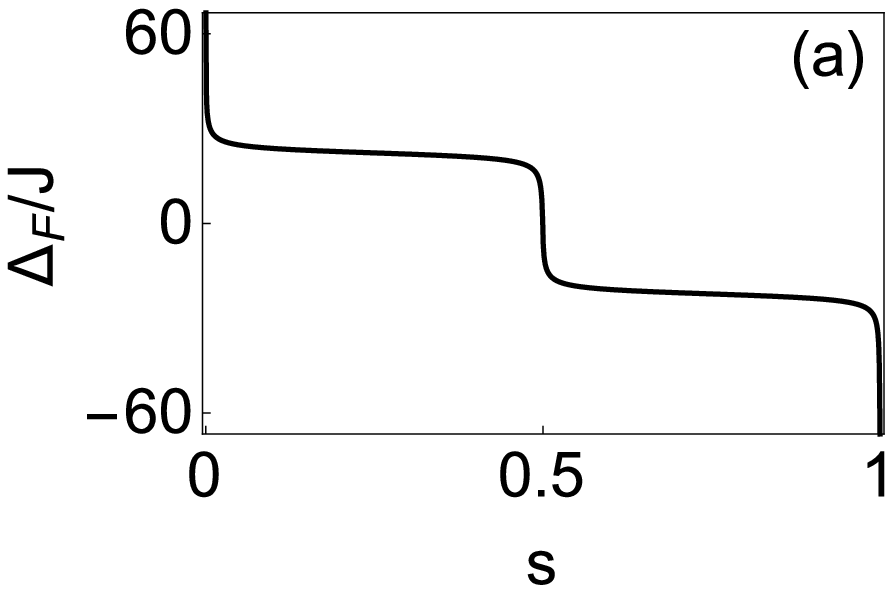}
\includegraphics[height=2.86cm,angle=0]{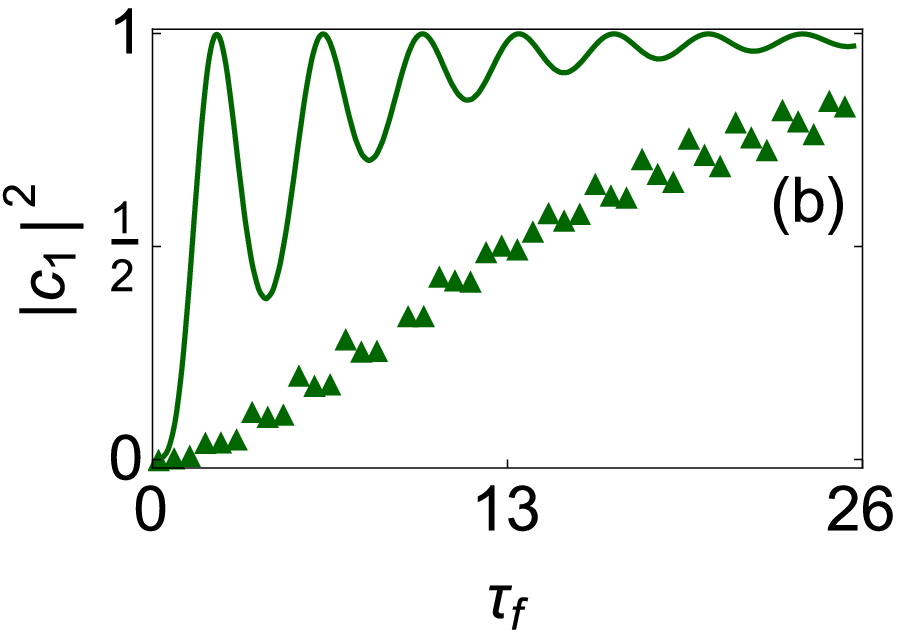}
\end{center}
\caption{\label{cotu}
(Color online). (a) Time dependence of the bias with FAQUAD. 
(b) $|c_1|^2$ vs. $\tau_f$ for linear-in-time bias (green triangles) and FAQUAD (solid green line). 
$\Delta(0)/J=66.7$, $U/J=22.3$ and $\tau_f=Jt_f/\hbar$.}
\end{figure}
%
%
%
%
%
%
The results of FAQUAD and the linear protocol are compared in Fig. \ref{levels_ap} (b). The probability of the first peak for FAQUAD, 0.998
at $\tau_f=1.2$, is achieved with the linear ramp for $\tau_f=43$.  

-Cotunneling. In a speeded-up cotunneling shown in Fig. \ref{tunneling} (b) the goal is to 
drive the system fast from $|\phi_1(0)\rangle=|0,2\rangle$ to $|\phi_1(t_f)\rangle=|2,0\rangle$ intermediated by $|1,1\ra$ 
(the Hamiltonian  $H_0$
does not  connect $|2,0\rangle$ and $|0,2\rangle$ directly). We impose $\Delta(0)\gg U,J$ and $\Delta(t_f) =-\Delta(0)$ to have $|0,2\rangle$ and $|2,0\rangle$ as the ground state at initial and final times  respectively.
The energy levels versus $\Delta$ are depicted in  
Fig. \ref{levels_ap} (a) for repulsive interaction ($U>0$). 
Figure \ref{cotu} (a) shows the FAQUAD trajectory for $\Delta(t)$
for the repulsive strong-interaction regime, $U/J=22.3$.
Fig. \ref{cotu} (b) depicts the final probabilities of the bare state $|2,0\ra$ for FAQUAD and a linear protocol
that needs about $\tau_f=65$ to achieve the value of the first peak of the FAQUAD method
($|c_1|^2=0.998$ at $\tau_f=2.3$).  
The minima in the FAQUAD probability go in this case below the lower envelope $1-4\tilde{c}^2/t_f^2$
predicted by perturbation theory. The reason is a leak through the narrow avoided crossing at 
$\Delta=0$ from the second to the third energy level, 
see Fig. \ref{levels_ap} (a). The leak occurs at total process times in which the first avoided crossing produces a minimum 
of the ground state probability.          

{\it Collective superpositions of rotating and non-rotating atoms on a ring.} Creating a macroscopic or mesoscopic superposition of a many-particle system is a difficult task and of interest for research in quantum information, quantum metrology, and fundamental aspects of quantum mechanics. However, it was recently proposed that a low-dimensional gas of interacting bosons in the Tonks-Girardeau (TG) limit \cite{Girardeau:60} placed on a ring can be perturbed in such a way, that a robust superposition of two angular momentum states can be achieved. This perturbation corresponds to the introduction of a narrow potential, which is then accelerated to a certain value to spin up the gas \cite{Hallwood:10}.

\begin{figure}[tb]
  \begin{center}
    \includegraphics[height=2.95cm,angle=0]{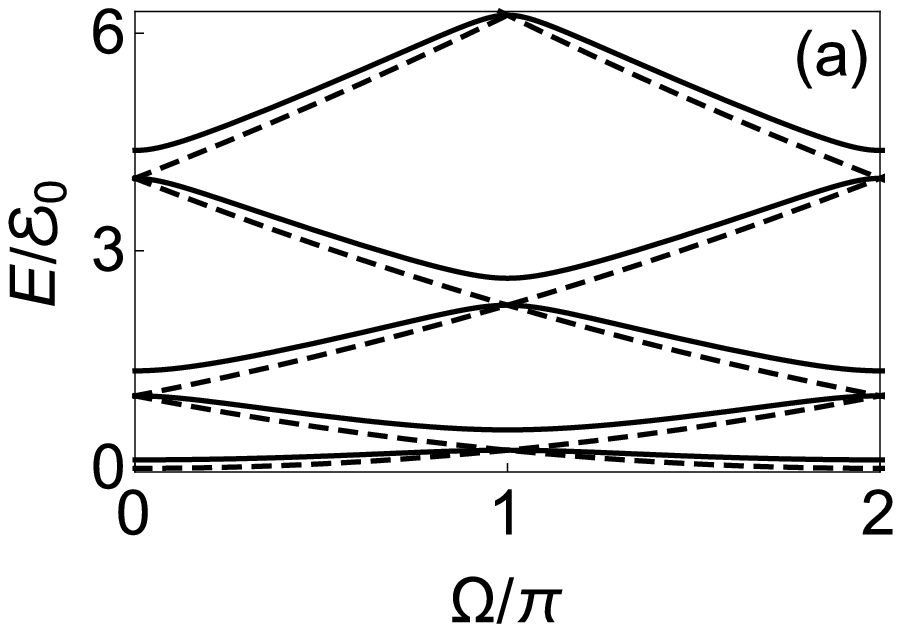}
    \includegraphics[height=2.95cm,angle=0]{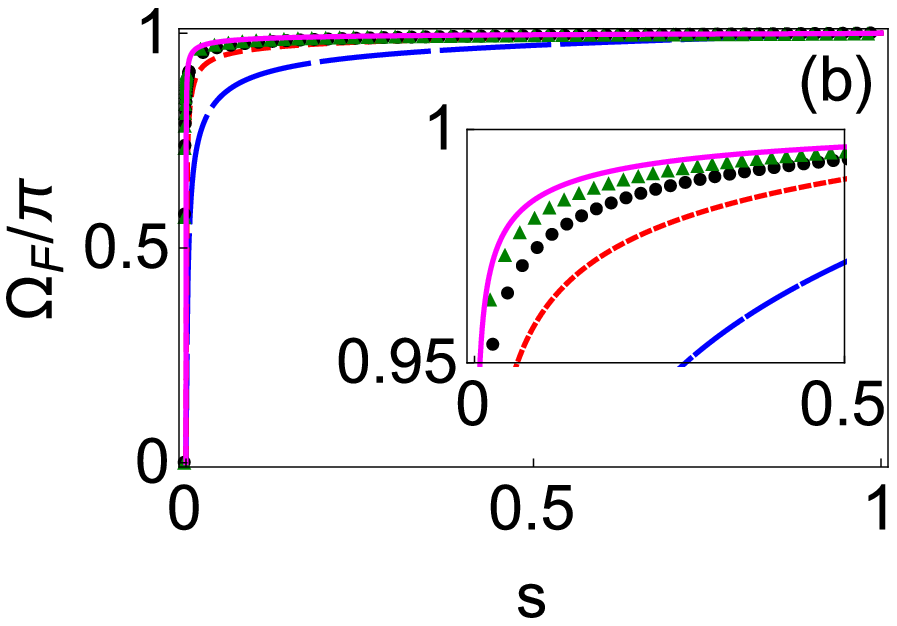}
  \end{center}
  \caption{\label{spoon}
  (Color online) (a) Single-particle energy levels for $U_0=0$ (dashed lines) and $U_0ML/\hbar^2=4$ (solid lines) in units of ${\cal{E}}_0=2\pi^2\hbar^2/(ML^2)$. The ordering is $E_1(n=0)<E_2(n=1)<E_3(n=-1)<E_4(n=2)<E_5(n=-2)<...$. (b) $\Omega_F(s)$ for $N=1,3,5,7,9$, from the bottom up to the top.}
\end{figure}

For a single particle this is described by
$
i\hbar \partial_t \psi (x,t)\!=\!\big\{-\frac{\hbar^2}{2M}\frac{\partial^2}{\partial x^2}\!+\!U_0\delta[x-x_0(t)]\big\} \psi(x,t),
$
where the stirrer is represented by a $\delta$-function of strength $U_0$ and periodic BC are assumed. In a co-moving frame one can then define $y=x-x_0(t)$ and the Hamiltonian is $H=\frac{1}{2M} \big[\hat P_y - \hbar\Omega(t)/L \big]^2+U_0\delta(y)$, where $L$ is the ring perimeter, $\hbar\Omega=M\dot x_0$, and $\hat P_y=-i\hbar\partial/\partial y$. The energy eigenvalues are 
$
E(n)=\frac{2\hbar^2\pi^2}{L^2M}\alpha_n^2,
$
and the $\alpha_n$ are solutions of 
$
\frac{4\pi\hbar^2\alpha_n}{MLU_0}=\cot(\pi \alpha_n- \Omega/2)+ \cot(\pi \alpha_n+\Omega/2).
$
For $U_0\rightarrow 0$, the $\alpha_n$ tend to $n-\Omega/(2\pi)$, with $n=0,\pm1,\pm2,\dots$, 
where the different signs are for clockwise or counter-clockwise rotation in the lab frame, and the $n$-th eigenstates are plane waves with momentum $n\hbar 2\pi/L$. For $0<\Omega<\pi$ the energies in the moving frame increase for $n\le 0$ and decrease for $n>0$. For $U_0=0$ the spectrum shows degeneracies at $\Omega=0,\pi$, which turn into avoided crossings once the stirrer couples different angular momentum eigenstates, as shown in Fig.~\ref{spoon}(a). Adiabatically increasing the stirring frequency from $\Omega=0$ to $\pi$ then allows to drive the system into a superposition of two angular momentum states and for a TG gas with an odd number of particles $N$ it can be shown that the ground state at $\Omega=\pi$ corresponds to macroscopic superposition between the angular momentum states $0$ and $N\hbar$.

\begin{figure}[t]
  \begin{center}
    \includegraphics[height=2.89cm,angle=0]{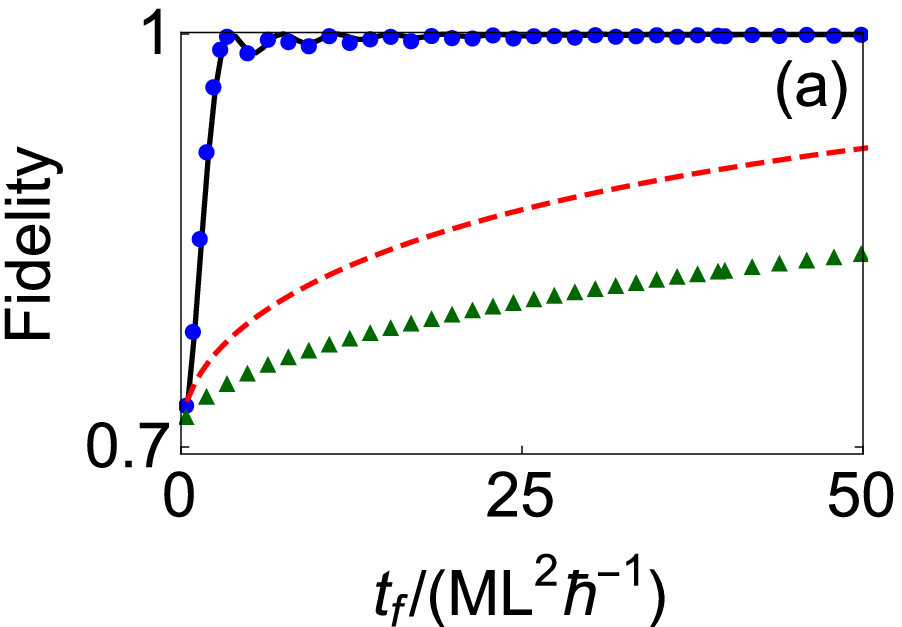}
    \includegraphics[height=2.89cm,angle=0]{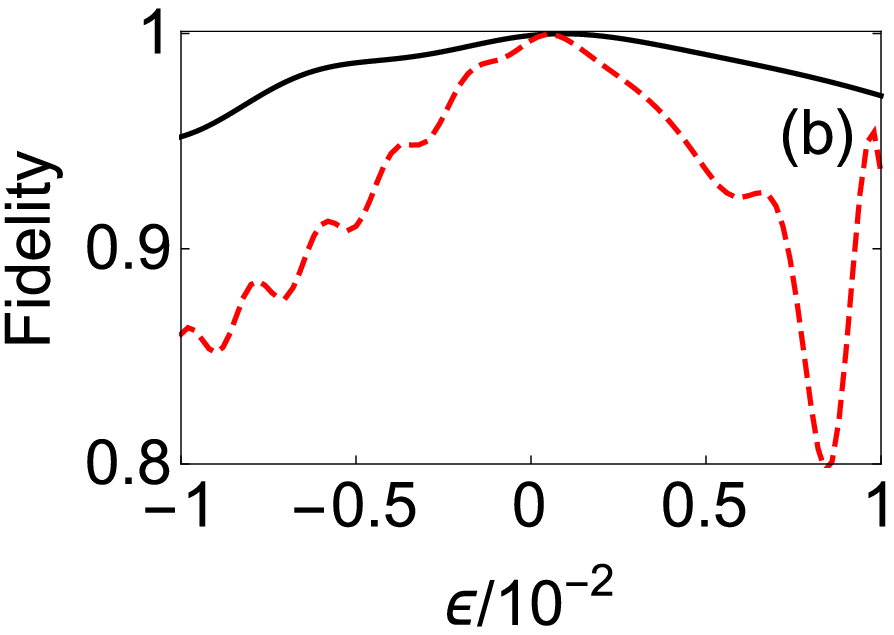}
  \end{center}
  \caption{\label{spoon2}
 (Color online). (a) Fidelity $|\la\Psi_{TG}(t_f)|\Phi_{TG}\ra|$  for $N=3$ (FAQUAD (solid black line) and linear $\Omega(t)$ (short-dashed red line)) and $N=9$ (FAQUAD (blue circles) and linear $\Omega(t)$ (green triangles)).  $\Psi_{TG}(t_f)$ is the time-evolved TG state starting from the ground state for $\Omega=0$, and $\Phi_{TG}$ is the ground state of the TG gas at $\Omega=\pi$. (b) Fidelity $|\la\Psi_{TG}(t_f)|\Phi_{TG}\ra|$ versus $\epsilon$ if FAQUAD is applied following a {\it wrong} $\Omega_e(t)=\Omega_F(t)(1+\epsilon)$ for $N=3$ (solid black line) and $N=9$ (short-dashed red line). Here $U_0ML/\hbar^2=0.5$.}
\end{figure}

To design an optimal $\Omega(t)$ for the TG gas, we first note that the fidelity depends mostly on leakage from the highest occupied levels. This can be understood by invoking the Bose-Fermi mapping theorem and realizing that the fermionic character of the particles prevents any low lying states from making transitions.  We can therefore optimize  $\Omega_F(s)$ for the avoided crossing of the highest occupied level as shown in  Fig.~\ref{spoon}(b). The corresponding final state fidelities for $N=3$ and $N=9$ with respect to the exact ground states clearly outperform the ones for the linear ramp, see Fig.~\ref{spoon2}(a). The linear ramp fidelity deteriorates as $N$ increases whereas, remarkably, the fidelity of the FAQUAD protocol stays constant. The effect of an error of the form $\Omega_e(t)=\Omega_F(t)(1+\epsilon)$ is shown in Fig.~\ref{spoon2}(b). 
%
%
%
%
%
%
%
%
%
%
%
%
%
%

{\it{Discussion}.} 
The FAQUAD approach to speed up adiabatic manipulations of quantum systems achieves significant time shortenings by distributing homogeneously the
adiabaticity parameter along the process time while satisfying the boundary conditions of the
control parameter.
We have derived general time scales and    
we have demonstrated its applicability in  different systems, in particular where 
other approaches are not available,  and expect a broad range of applications, in quantum, optical and mechanical
systems,  due to the ubiquity of adiabatic methods. 

A natural extension is to attempt a scheme similar to Eq. (\ref{f_adiabatic}) in a superadiabatic 
rather than an adiabatic frame \cite{Sara}. However to produce a STA the superadiabatic states must coincide with eigenstates of the Hamiltonian at boundary times. This implies additional boundary conditions on the control parameter \cite{Sara}. 
For example, in the two-level model, $\dot{\Delta}(0)=\dot{\Delta}(t_f)=0$, but the equation substituting Eq. (\ref{f_adiabatic}) 
in the lowest superadiabatic scheme beyond the adiabatic level, is a second order differential equation for $\Delta$. We would then have four conditions (on $\Delta$ and $\dot{\Delta}$) that cannot be satisfied with two integration constants plus the $c$. Since the same problem appears in higher superadiabatic orders, the adiabatic frame of FAQUAD is in fact optimal in the series of iterative superadiabatic frames.       
\begin{acknowledgments}
We thank M. Palmero, X. Chen, and S. Ib\'a\~nez for discussions.   
Support by 
the Basque Country Government (Grant No.
IT472-10),
Ministerio de Econom\'{i}a y Competitividad (Grant No.
FIS2012-36673-C03-01),  program UFI 11/55 of UPV/EHU, and the Okinawa 
Institute of Science and Technology Graduate University is acknowledged. 
This publication has emanated from research conducted with the financial support of Science Foundation
Ireland under the International Strategic Cooperation Award Grant Number SFI/13/ISCA/2845.
S. M.-G. acknowledges a fellowship by UPV/EHU.  
\end{acknowledgments}
%

\end{document}